# Network Simulator – Visão Geral da Ferramenta de Simulação de Redes

Marcos Portnoi[*]

Rafael Gonçalves Bezerra de Araújo[**]

Orientador: Prof. Sérgio de Figueiredo Brito[***]

**Resumo**

*Este artigo visa descrever a ferramenta de simulação de redes de computadores denominada Network Simulator – NS. Uuma visão geral de sua filosofia será oferecida, abordando também seu funcionamento e características. Ao final, serão apresentandas todas as etapas de preparação para simulação de um modelo simples no NS.*

**Introdução**

A pesquisa científica ou a administração de sistemas freqüentemente avaliam propostas para melhoria de processos vigentes ou para solução de problemas. A Avaliação de Desempenho de Sistemas oferece técnicas para abordar essa questão [1][2]. Uma destas abordagens consiste em **modelar** um determinado **sistema**, contendo as variáveis de interesse, de forma que este modelo possa ser tratado matematicamente. O tratamento do modelo oferecerá diversos resultados de acordo com as situações desejadas. As variáveis de interesse num sistema de redes de computadores podem ser: o atraso fim-a-fim, pacotes perdidos, vazão, dentre outras.

A solução do modelo pode ser desenvolvida através de métodos analíticos ou através de simulação. O método analítico, ou algébrico, implica em posse de profundo conhecimento matemático, mas confere exatidão. Dependendo do modelo matemático, a resolução torna-se extenuante, o que obriga a simplificações do modelo, o que pode resultar em imperfeições na representação do sistema. O método por simulação, por outro lado, permite a confecção de modelos complexos e resolução destes com menor desenvolvimento matemático. Para isso, emprega-se poder computacional para as iterações numéricas requeridas implicando, a depender dos resultados desejados, grande consumo computacional.

---

[*] Graduando em Engenharia Elétrica pela Universidade Salvador – UNIFACS, membro do IEEE – Institute of Electrical and Electronics Engineers, Gestor com mais de 12 anos de experiência em finanças, administrativo e comercial, Pesquisador em Iniciação Científica em Redes de Computadores no Núcleo de Pesquisa em Redes de Computadores – NUPERC – da UNIFACS.
[**] Graduando em Engenharia Elétrica pela UNIFACS, Pesquisador em Iniciação Científica em Redes de Computadores no NUPERC.
[***] Doutorando em Ciência da Computação pela Universidade Federal de Pernambuco, Mestre em Engenharia Elétrica pela Universidade Federal da Paraíba, Coordenador do Curso de Engenharia da UNIFACS, Professor do Depto. de Engenharia e Arquitetura e do Mestrado em Redes de Computadores e Pesquisador do NUPERC.

Um dos simuladores que tem sido utilizado com grande freqüência em pesquisas em redes de computadores é o *Network Simulator* – ns – [3]. Concebido em 1989 a partir de uma variação do REAL Network Simulator [4], um projeto da Cornell University, EUA, o *ns* tem evoluído desde então, sempre com suporte e apoio de várias organizações durante períodos. Atualmente o desenvolvimento do *ns* é suportado pelo DARPA (*Defense Advanced Research Projects Agency*, EUA) através do projeto SAMAN [5] e pela NSF (*National Science Foundation,* EUA) através do projeto CONSER [6], em colaboração com outros pesquisadores como o centro ICIR [7]. O simulador já recebeu apoio do Lawrence Berkeley National Laboratory, do Xerox PARC (Palo Alto Research Center), da Universidade da Califórnia em Berkeley, Sun Microsystems e também agrega diversos módulos contribuídos por pesquisadores independentes. É um software de código livre e fornecido gratuitamente [3]. Uma lista de discussão é mantida pelos desenvolvedores, onde os pesquisadores de diversas partes do mundo podem trocar idéias e experiências, e também propor correções para o código do simulador, que após avaliadas podem ser incorporadas. Estes pesquisadores, oriundos de países como Estados Unidos, Índia, Inglaterra, Itália, Taiwan e também Brasil, contribuem para o valor desta ferramenta.

O *ns* é um simulador de eventos discretos, focado para o desenvolvimento de pesquisas em redes de computadores. Ele prevê suporte a TCP e variantes do protocolo (Tahoe, Reno, New Reno, Vegas, etc.), multicast, redes sem fio (*wireless*), roteamento e satélite. Implementa filas de roteamento tipo *droptail*, Diffserv RED [8], *fair queueing* (FQ), *stochastic fair queueing* (SFQ), *class-based queueing* (CBQ), dentre outras [9]. Tem facilidades de *tracing*, que é a coleta e registro de dados de cada evento da simulação para análise posterior. Possui um visualizador gráfico para animações da simulação (*nam – network animator*), *timers* e escalonadores, modelos para controle de erros e algumas ferramentas matemáticas como gerador de números aleatórios e integrais para cálculos estatísticos. Inclui também uma ferramenta de plotagem, o *xgraph*, e vários tipos de geradores de tráfego.

O *ns* foi desenvolvido na linguagem orientada a objetos C++, de forma modular. O uso desta linguagem nos módulos confere velocidade e mais praticidade na implementação de protocolos e modificação de classes. A interface com o usuário, configuração, estabelecimento de parâmetros e manipulação de objetos e classes é feita em modo texto, através da linguagem interpretada OTcl [10], que também é orientada a objetos.

## Instalação do *ns*

O *ns* foi construído para rodar preferencialmente em plataformas Unix (FreeBSD, SunOS, Solaris, Linux, dentre outras). Pode também rodar em plataforma

Microsoft Windows, apesar de que sua instalação não é tão automatizada e não segue os padrões de instalação dos software feitos para Windows. Este artigo descreverá características do simulador em sistema operacional Linux (distribuição RedHat 7.1) e arquitetura Intel PC.

Fornecido em fontes, o *ns* é compilado durante o processo da instalação. Assim, faz-se necessário um compilador C++ no computador onde será instalado. Outros pré-requisitos para o ambiente são fornecidos em [3]. Como o simulador é compilado na instalação, ele pode, a princípio, ser executado em qualquer arquitetura de computador, como Intel/AMD ou RISC.

O pacote do simulador é composto dos seguintes módulos básicos:

- Tcl/Tk: Interpretador de linguagem Tcl, que é a interface do simulador com o usuário.
- OTcl: suplemento de orientação a objetos para o Tcl.
- Tclcl: implementação de classes para Tcl.
- ns-2: classes do simulador propriamente dito.
- nam-1: visualizador e animador gráfico de topologias de rede e simulação.
- xgraph: ferramenta de plotagem de gráficos.
- cweb e SGB: bibliotecas requeridas para sgb2-ns e gt-itm, abaixo.
- Gt-itm, gt-itm e sgb2-ns: gerador de topologias.
- zlib: ferramenta de compressão de arquivos.

Há duas formas de instalar o simulador: através do *ns-allinone*, que reúne todos os pacotes acima descritos, ou obtendo cada pacote e fazendo sua instalação separadamente. Nem todos os módulos acima são requeridos para que o simulador funcione, portanto pode-se preparar ambientes com instalações mínimas e outros com mais ferramentas. Já outros módulos externos (como *perl*) são necessários para realizar certas funções. Em [3] mais detalhes estão disponíveis.

Com o pacote *ns-allinone*, a instalação é razoavelmente simples, desde que o ambiente tenha todos os pacotes externos necessários. Deve ser descompactado com auxílio de uma ferramenta apropriada (como o *ark*) para o diretório onde se deseja que o simulador permaneça (por padrão, recomenda-se colocar em /usr/local). Dentro deste diretório, executa-se o *script* de instalação: *./install.*

Os fontes de cada pacote do simulador serão então apropriadamente compilados na ordem correta (processo que pode ser bastante moroso em computadores mais antigos). Se não houver nenhum erro na compilação, o *script* dará mensagem de sucesso e fornecerá instruções complementares para modificação de variáveis de ambiente do sistema operacional, como por exemplo a variável PATH. Essas modificações têm de ser feitas manualmente e para cada usuário que se utilizará do simulador, no arquivo que contém as configurações do usuário (em RedHat Linux, o arquivo oculto .bash_profile, que fica nos diretórios individuais de cada usuário). Se houver alguma mensagem de erro, o operador deve procurar auxílio na página de apoio [12] ou ainda na lista de discussão mantida pelos desenvolvedores.

Uma vez terminada a compilação, deve-se proceder à validação do simulador, que consiste em executar várias simulações pré-programadas e comparar seus resultados contra resultados-padrão. Isso é feito através do comando *./validate*, dentro do diretório ns (no caso da versão 9, ns-2.1b9). Dependendo da versão do Unix utilizada e da arquitetura do computador, um ou mais testes podem diferir dos resultados-padrão. Novamente, o operador deve pesquisar as razões destas diferenças na Internet (nas páginas de apoio ou listas de discussão) de forma a verificar se tais incongruências prejudicarão resultados posteriores (há diferenças causadas apenas por reordenamento de dados, o que não invalida os resultados).

## Estrutura

A interface entre o usuário e o *ns* dá-se através da linguagem *script* OTcl. Segundo os desenvolvedores, a divisão em duas linguagens (OTcl e C++) objetiva dar ao simulador tanto velocidade e poder, quanto flexibilidade e facilidade de mudança de parâmetros. O núcleo do simulador é escrito em C++, conferindo velocidade, mas esta linguagem torna-se lenta para manipulação constante ou mudança de parâmetros. OTcl, por ser interpretada, é bem mais lenta, porém pode ser facilmente alterada. Além do mais, os objetos compilados são disponibilizados para o interpretador OTcl por *linkagem*, o que virtualmente cria um objeto OTcl para cada objeto C++, e que podem ser manipulados através das facilidades da OTcl. A Figura 1 mostra a construção geral do *ns* (algumas figuras neste artigo foram adaptadas de [13]). Um usuário comum atua no perímetro "*tcl*", escrevendo *scripts* em OTcl e executando simulações. Os escalonadores de eventos e os componentes de rede são implementados em C++ e disponibilizados ao interpretador OTcl através de uma replicação feita pela camada *tclcl,* que recria os objetos C++ em objetos OTcl, e que podem finalmente ser manipulados por esta última (processo denominado *linkage*). Todo o conjunto constitui-se no *ns*, que é um interpretador de OTcl com bibliotecas de simulação para redes de computadores.

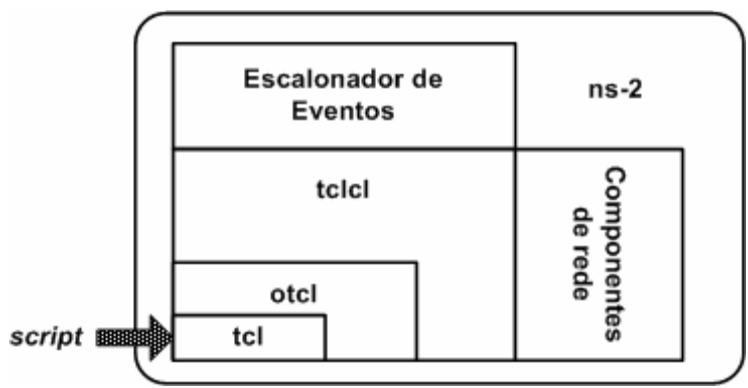

**Figura 1: Arquitetura do *ns*.**

A Figura 2 mostra a árvore parcial de hierarquia de classes do ns. Os objetos de rede são criados segundo essas classes.

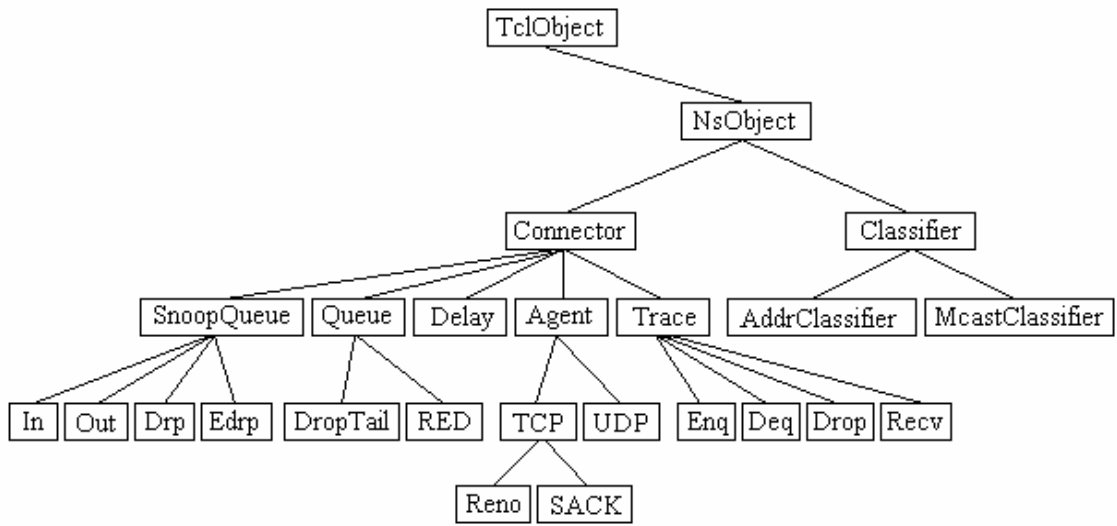

**Figura 2: Hierarquia de classes parcial.**

## Como Utilizar: a Interface

A operação básica do ns está ilustrada na Figura 3.

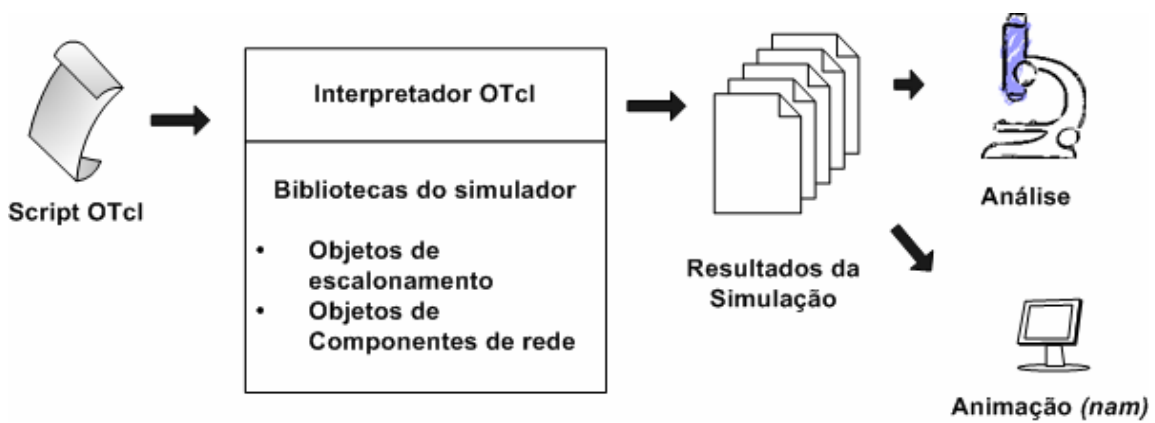

**Figura 3: Esquema de utilização do *ns*.**

Para montar uma simulação no *ns* é preciso então, primeiramente, escrever um *script* em OTcl. Este *script* contém as seguintes partes básicas:

- criação do objeto Simulador

- abertura de arquivos para *tracing* e análise posterior
- criação da topologia de rede
    - criação de nós ou nodos
    - conexão dos nós entre si (*links*)
    - criação das filas de saída
- criação dos agentes de 4ª. camada e conexão com nós
- criação dos geradores de tráfego (nível de Aplicação) e conexão com agentes de 4ª. camada (nível de Transporte)
- programação dos escalonadores e *timers*
- fechamento da simulação, animação e geração de estatísticas

O processo de simulação pode ser assim resumido:

- confecção do *script* (arquivo texto comum)
- execução do *script* com o comando *ns nomedoscript.tcl*
    - arquivos de *tracing* serão gerados com registro de cada evento simulado
- após conclusão da simulação:
    - imprimir estatísticas calculadas no *script*
    - visualizar os eventos com o *nam*
    - analisar resultados através dos arquivos de *tracing* com apoio de ferramentas apropriadas (*awk*)

Um ponto importante a observar é que o *ns* não fornece estatísticas de simulação de modo automático; estas devem ser obtidas através de procedimentos matemáticos no *script* ou pela manipulação de objetos especiais chamados *monitores*. Pode-se, ainda, usar ferramentas para análise dos arquivos de *tracing* gerados durante a simulação, que são os verdadeiros resultados da simulação, conforme a Figura 3. Estes arquivos, com formatação específica, registram cada evento gerado pelos escalonadores. As ferramentas para análise dos arquivos de *tracing* devem então ser capazes de ler os dados gravados nestes arquivos e efetuar os cálculos desejados. Uma destas ferramentas, muito utilizada, é o *awk* [11], uma linguagem desenhada para buscar padrões dentro de um arquivo e efetuar ações programadas. O animador *nam* pode também ser usado para analisar visualmente a simulação e obter algumas estatísticas, mas ele não é apropriado para análises mais profundas. Se nada for feito, o simulador apenas rodará o *script*, gerará os arquivos de saída (*tracing*) e encerrará, sem nada mostrar ao usuário. Em primeiro momento, esta característica um tanto não-amigável e não-imediatista do *ns* pode frustrar o usuário iniciante. Entretanto, essa é a realidade de bom número dos simuladores existentes na categoria do *ns*, que não foram feitos com ótica didática.

## Simulação de Um Modelo

O modelo proposto para simulação, apresentado na topologia ilustrada na Figura 4, contém dois geradores de tráfego, um do tipo CBR (*Constant Bit Rate*) [9] e outro do tipo exponencial *on-off*. Todos os nós têm filas de saída tipo *droptail*, com exceção do nó 2, que recebe uma fila tipo *Stochastic Fair Queueing* (SFQ) [9]. O nó 3 agirá como sorvedouro, recebendo o tráfego dos geradores. O *script* completo está disponibilizado em [14]. O gerador exponencial estará no nó 0 e enviará seus dados para o nó 3, através do nó 2. O gerador CBR estará no nó 1 e enviará para o nó 3, também através do nó 2.

O primeiro passo é a criação do objeto simulador, que é feita da seguinte maneira:

*set ns [new Simulator]*

No modelo proposto foi estipulado o tempo máximo de simulação:

*set MAX_TIME 500*

Após a criação do objeto simulador, abre-se os arquivos *tracing* para posterior análise:

*set nf [open out.nam w]*
*set tr [open out.tr w]*
*$ns trace-all $tr*
*$ns namtrace-all $nf*

O primeiro arquivo (*out.nam*) conterá o registro de eventos simulados no formato para leitura do animador *nam*, e o segundo arquivo (*out.tr*) conterá os registros dos eventos de maneira mais completa, permitindo análises profundas da simulação.

A topologia da rede é criada da seguinte maneira:

*set n0 [$ns node]*
*set n1 [$ns node]*
*set n2 [$ns node]*
*set n3 [$ns node]*
*$ns duplex-link $n0 $n2 10Mb 10ms DropTail*
*$ns duplex-link $n1 $n2 10Mb 10ms DropTail*
*$ns duplex-link $n2 $n3 10Mb 10ms SFQ*

Nas quatro primeiras linhas do código acima são criados os nós. Configura-se então a ligação entre estes nós, com *links* do tipo *duplex*, cada um com largura de banda de 10Mbits e atraso de propagação de 10 ms.

A *procedure* abaixo cria uma 4ª. camada do tipo UDP, um gerador de tráfego exponencial *on-off* e faz a conexão de ambos, entre os nós 0 e 3 (sorvedouro). O manual do *ns* disponível em [3] traz mais detalhes sobre a configuração dos geradores de tráfego. A *procedure* também define uma cor para uso do *nam*.

```
proc attach-expoo-traffic {node sink size burst idle rate class color} {
    set ns [Simulator instance]
    set source [new Agent/UDP]
    $ns attach-agent $node $source
    $source set class_ $class
    $ns color $class $color
    set traffic [new Application/Traffic/Exponential]
    $traffic set packetSize_ $size
    $traffic set burst_time_ $burst
    $traffic set idle_rate_ $idle
    $traffic set rate_ $rate
    $traffic attach-agent $source
    $ns connect $source $sink
    return $traffic
}
set traffgen0 [attach-expoo-traffic $n0 $sink0 1000 800ms 2ms 5M 1 Green]
```

Para o nó 1, as linhas abaixo criam a 4ª. camada tipo UDP e um gerador de tráfego CBR, e também a conexão destes entre os nós 1 e 3.

```
set udp1 [new Agent/UDP]
$udp1 set class_ 2
$ns attach-agent $n1 $udp1
set cbr0 [new Application/Traffic/CBR]
$cbr0 set packetSize_ 1000
$cbr0 set interval_ 0.005
$cbr0 attach-agent $udp1
$ns connect $udp1 $sink0
```

As linhas abaixo trazem a programação do escalonador de eventos:

```
$ns at 0.0 "$traffgen0 start"
$ns at 1.0 "$cbr0 start"
$ns at [expr $MAX_TIME-1] "$traffgen0 stop"
$ns at [expr $MAX_TIME-1] "$cbr0 stop"
$ns at $MAX_TIME "finish"
```

As primeiras quatro linhas definem em que intervalos de tempo cada gerador de tráfego deve iniciar e parar suas atividades. A expressão *[expr $MAX_TIME-1]* significa o tempo total da simulação menos um segundo, que é quando o tráfego interrompe suas atividades.

O código abaixo define o processo de finalização da simulação, chamado pelo escalonador. Os arquivos de *trace* são fechados e mostra-se algumas estatísticas coletadas durante a simulação (através do objeto *LossMonitor*). O visualizador *nam* também é executado aqui. A Figura 4 traz o *nam* em ação.

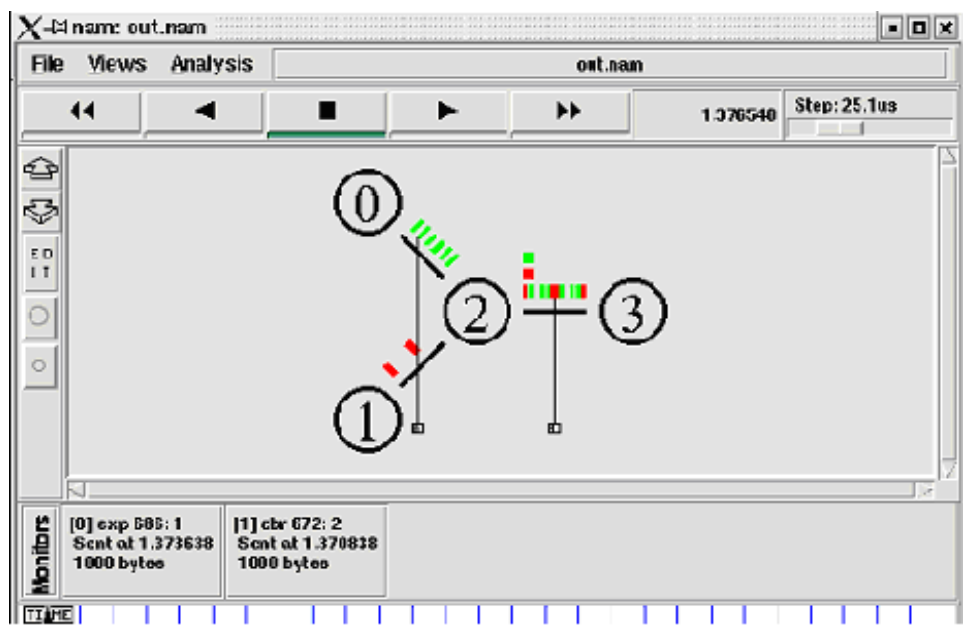

**Figura 4: Tela do *nam* executando a simulação.**

```
proc finish {} {
global ns nf tr MAX_TIME sink0
$ns flush-trace
close $nf
close $tr
set now [$ns now]
puts "Estatisticas:"
puts "Tempo Simulacao:  $now s"
puts "Pacotes recebidos no nodo 3:  [$sink0 set npkts_]"
puts "Bytes recebidos no nodo 3:  [$sink0 set bytes_]"
puts "Utilizacao do link:  [expr [$sink0 set bytes_ ]*8. / (10000000. * $now) *100.]%"
exec nam out.nam &
exit 0
}
```

A ultima linha do script, finalmente, inicia a execução do simulador.

*$ns run*

Para este modelo e parâmetros utilizados, o *script* forneceu os seguintes resultados:

*Tempo Simulacao: 500 s*
*Pacotes recebidos no nodo 3: 384377*
*Bytes recebidos no nodo 3: 334576500*
*Utilizacao do link: 53.532239999999994%*

## Considerações Finais

O tema Rede de Computadores possui um imenso leque de oportunidades de pesquisa, seja em Qualidade de Serviços, seja em redes sem fio (*wireless*), seja em integração de serviços de voz e vídeo em redes IP. A utilização de simuladores para testar e validar os modelos propostos é imprescindível em todas essas áreas.

Atualmente, o *ns* é manuseado por vários grupos de pesquisa atuando nos campos acima e também em redes sem-fio celulares, satélite e em simulação de ataques DoS (*Denial of Service*) para testes de segurança. Esses grupos contribuem para manter vivo o projeto do *ns*, atualizando-o com novos protocolos e suporte a novas tecnologias e maneira rápida e colaborativa. A adoção do *ns* garante o uso de uma ferramenta de ponta pelo pesquisador em seus trabalhos, podendo influir na qualidade de seus resultados e em sua velocidade.

## Referências Bibliográficas